\documentclass[12pt,superscriptaddress,aps,prd,preprint]{revtex4}
\usepackage{amsmath}
\usepackage{amssymb}
\makeatletter
\newcommand{\bea}{\begin{eqnarray}}
\newcommand{\eea}{\end{eqnarray}}

\begin{document}
\title{Ricci dark energy in Chern-Simons modified gravity}
\author{J. G. Silva, A. F. Santos}
\affiliation{Instituto de F\'{\i}sica, Universidade Federal de Mato Grosso,\\
78060-900, Cuiab\'{a}, Mato Grosso, Brazil}
\email{jucelia,alesandroferreira@fisica.ufmt.br}

\begin{abstract}
In this work, we have considered the Ricci dark energy model, where the energy density of the universe is proportional to the Ricci scalar curvature, in the dynamic Chern-Simons modified gravity. We show that in this context the evolution of the scale factor is similar to that displayed by the modified Chaplygin gas.
\end{abstract}

\maketitle
Currently the accelerated expansion of the universe has been strongly confirmed by some independent experiments such as the Cosmic Microwave Background Radiation (CMBR) \cite{Spergel}  and Sloan Digital Sky Survey (SDSS) \cite{Adelman}. In an attempt to explain this phenomenon there are two possible paths; first option -  propose corrections to general relativity, second option - assuming that there is a dominant component of the universe, a kind of antigravity called dark energy. Any way that we intend to follow, there are numerous models that attempt to explain this effect. 

One of the models of modified gravity that has stood out in recent years is the Chern-Simons modified gravity, which was initially developed in \cite{Jackiw}. This modification consists in the addition of the Pontryagin density, which displays violation of parity symmetry in Einstein-Hilbert action. In four dimensions the Pontryagin density is simply a topological term, unless the coupling constant is not constant or promoted to a scalar field. The Chern-Simons modification is not a random extension, but it is actually motivated by both string theory, as a ne\-cessary anomaly-canceling term to conserve unitarity \cite{Polchinski}, and loop quantum gravity \cite{Ashtekar}. For further details this proposal for a correction to general relativity, see the revision made ​​in \cite{Yunes}. In this paper, we will study the dynamic formulation of Chern-Simons gravity, where the coupling constant is promoted to a scalar field, recent studies in this formulation include: neutron star binary \cite{Yagi}, spacetime with and without black holes \cite{Tanabe}, among others.

From among the various models proposed for dark energy there are some that are based on the holographic principle, known as holographic dark energy \cite{Li} \cite{Hsu}. Such models are based on the idea that the energy density of a given system is proportional to the inverse square of some characteristic length of the system, for example, in \cite{Cohen} energy density is proportional to the Hubble scale, the future event horizon is used as the characteristic length in \cite{Li}. However, these models exhibit some fundamental problems such as the problem of fine tuning and/or coincidence problem. From these studies, it has been proposed in \cite{Gao} a model of dark energy where the characteristic length is given by the average radius of the Ricci scalar, $R^{-1/2}$. Thus, the dark energy density is proportional to the Ricci scalar, i.e.,
\bea
\rho_x=-\frac{\alpha}{16 \pi}R,
\eea
where $R$ is the Ricci scalar curvature and $\alpha$ is a constant to be determined. This model avoids the problems of fine tuning and coincidence that arise in previous models. It is a phenomenologically viable model  and displays results similar to that presented by the cosmological model $\Lambda CDM$. In this work we will dedicate ourselves to investigate the Ricci dark energy in the context of Chern-Simons modified gravity for homogeneous and isotropic universe described by the Friedmann-Robertson-Walker (FRW) metric given by
\bea
ds^2=-dt^2+a(t)^2\left(\frac{dr^2}{1-kr^2}+r^2d\theta^2+r^2\sin^2\theta d\phi^2\right).
\eea
In this case the Ricci scalar curvature is given by
\bea
R=-6\left(\dot{H}+2H^2+\frac{k}{a^2}\right),
\eea 
where $k$ is the spatial curvature and $H=\frac{\dot{a}}{a}$ is the Hubble parameter. Then, the energy density becomes
\bea
\rho_x=\frac{3\alpha}{8\pi}\left(\dot{H}+2H^2+\frac{k}{a^2}\right). \label{density}
\eea
 
Now let's study this dark energy model in the context of the dynamics Chern-Simons modified gravity. The action that describes this theory is given by
\bea
S&=&\frac{1}{16\pi G}\int d^4x \Biggl[\sqrt{-g}R+\frac{l}{4}{\theta\, ^*RR}-\frac{1}{2}\partial^\mu\theta\partial_\mu\theta+ V(\theta)\Biggl]+ S_{mat.},
\eea
where ${^*RR}$ is the topological invariant called the Pontryagin term, $l$ is a coupling constant, the function $\theta$ is a dynamical variable, $S_{mat.}$ is the action of matter and $V(\theta)$ is the potential, that for simplicity we equal to zero. Varying the action with respect to the metric $g_{\mu\nu}$ and to the scalar field $\theta$ we obtain the field equations, respectively
\bea
G_{\mu\nu}+l C_{\mu\nu}=8\pi G T_{\mu\nu},\label{einstein}\\
g^{\mu\nu}\nabla_\mu\nabla_\nu\theta=-\frac{l}{64 \pi}{^*RR},\label{scalar}
\eea
where $G_{\mu\nu}$ is the Einstein tensor and $C_{\mu\nu}$ is Cotton tensor defined by
\bea
C^{\mu\nu}&=&-\frac{1}{2\sqrt{-g}}\Bigl[v_\sigma \epsilon^{\sigma\mu\alpha\beta}\nabla_\alpha R^\nu_\beta+\frac{1}{2}v_{\sigma\tau}\epsilon^{\sigma\nu\alpha\beta}R^{\tau\mu}\,_{\alpha\beta}\Bigl]+ (\mu\longleftrightarrow\nu), 
\eea
with $v_\sigma\equiv\nabla_\sigma \theta$, $v_{\sigma\tau}\equiv \nabla_\sigma \nabla_\tau\theta$. The energy-momentum tensor is divided into two parts
\bea
T_{\mu\nu}=T_{\mu\nu}^{RDE}+T_{\mu\nu}^{\theta},
\eea
where
\bea
T_{\mu\nu}^\theta=(\nabla_\mu\theta)(\nabla_\nu\theta)-\frac{1}{2}g_{\mu\nu}(\nabla_\lambda\theta)(\nabla^\lambda\theta)
\eea
is the energy-momentum tensor associated with the scalar field and 
\bea
T_{\mu\nu}^{RDE}=(\rho_x+p_x)U_\mu U_\nu+p_x g_{\mu\nu}
\eea
is the the energy-momentum tensor of the Ricci dark energy. We have that $U_\mu=(1,0,0,0)$ is the four-velocity, $\rho_x$ is the energy density (\ref{density}) and $p_x$ is the pressure of dark energy. 

Now let's study the component $00$ of equation (\ref{einstein}) which gives us the Friedmann equation
\bea
G_{00}+C_{00}=8\pi G (T_{00}^{RDE}+T_{00}^\theta),
\eea
where
\bea
G_{00}&=&3\left(\frac{\dot{a}^2}{a^2}+\frac{k}{a^2}\right),\\
T_{00}^\theta &=&\frac{1}{2}\dot{\theta}^2,\\
T_{00}^{RDE}&=&\rho_x=-\frac{\alpha}{16\pi}R,
\eea
and $C_{00}=0$. As discussed in \cite{Grumiller} the components of the Cotton tensor vanish for all spherically symmetric metrics. Thus the Friedmann equation becomes
\bea
\left(\frac{\dot{a}}{a}\right)^2+\frac{k}{a^2}=\alpha\left(\dot{H}+2H^2+\frac{k}{a^2}\right)+\frac{4\pi}{3}\dot{\theta}^2,
\eea
where we use natural units $c=G=1$. We have that 
$\dot{H}+2H^2=\frac{\ddot{a}}{a}+\left(\frac{\dot{a}}{a}\right)^2,$
thus
\bea
\alpha\frac{\ddot{a}}{a}+\left(\frac{\dot{a}}{a}\right)^2(\alpha-1)+\frac{k}{a^2}(\alpha-1)+\frac{4\pi}{3}\dot{\theta}^2=0.
\eea
Assuming a flat universe, $k=0$, we stay with
\bea
\alpha\frac{\ddot{a}}{a}+(\alpha-1)\left(\frac{\dot{a}}{a}\right)^2+\frac{4\pi}{3}\dot{\theta}^2=0.\label{meq1}
\eea

Now let's look at the field equation associated with the scalar field (\ref{scalar}). For the metric FRW we obtain that ${^*RR}=0$, thus this equation becomes
\bea
g^{\mu\nu}\nabla_\mu\nabla_\nu\theta=g^{\mu\nu}\left[\partial_\mu\partial_\nu\theta-
\Gamma_{\mu\nu}^{\lambda}\partial_{\lambda}\theta\right]=0.
\eea
Choosing $\theta=\theta(t)$, as was done in \cite{Jackiw}, this equation gives us
\bea
\ddot{\theta}+3\frac{\dot{a}}{a}\dot{\theta}=0.\label{meq2}
\eea

Therefore, our set of field equations consists of the equations (\ref{meq1}) and (\ref{meq2}). Now our goal is to resolve them to determine the evolution of the scale factor $a(t)$. From the equation (\ref{meq2}) we note that $\theta$ is related to $a(t)$ as
\bea
\dot{\theta}=C a^{-3},
\eea
where $C$ is a constant. Substituting this result into the equation (\ref{meq1}) we obtain that
\bea
\alpha\frac{\ddot{a}}{a}+(\alpha-1)\left(\frac{\dot{a}}{a}\right)^2+\beta a^{-6}=0,\label{meq3}
\eea
with $\beta\equiv\frac{4\pi C^2}{3}$.

For simplicity we make the following change of variable
\bea
u(a)=\frac{da}{dt} \,\,\,\,\,\,\, \Longrightarrow \,\,\,\,\,\,\, u\frac{du}{da}=\frac{d^2a}{dt^2}.
\eea
Thus the equation (\ref{meq3}) stays as
\bea
\alpha\, u\frac{du}{da}+(\alpha-1)\frac{u^2}{a}+\beta a^{-5}=0.
\eea
Solving this equation for $u(a)$ we find 
\bea
u(a)=\frac{\sqrt{\beta+a^{2+\frac{2}{\alpha}}(1+\alpha)c_1}}{a^2\sqrt{1+\alpha}},
\eea
where $c_1$ is an integration constant. Now returning to the original variables $a(t)$ and $t$ we can write
\bea
t=\sqrt{1+\alpha}\int\frac{a^2 da}{\sqrt{\beta+a^{2+\frac{2}{\alpha}}(1+\alpha)c_1}}.
\eea
Solving this integral for $a(t)$ we find that
\bea
t=\xi\, a^3\,_{2}F_{1}\left[\frac{1}{2},\frac{3\alpha}{2\zeta};1+\frac{3\alpha}{2\zeta};-\frac{a^{2+\frac{2}{\alpha}}c_1\zeta}{\beta}\right],
\eea
where $\,_{2}F_{1}$ is the hypergeometric function, $\zeta=1+\alpha$ and $\xi=\frac{1}{3}\sqrt{\frac{\zeta}{\beta}}$. At this time, let's analyze this result assuming $\alpha\approx\frac{1}{2}$ as found in \cite{Gao}. Thus
\bea
t=\frac{a^3}{\sqrt{6\beta}}\,_{2}F_{1}\left[\frac{1}{2},\frac{1}{2};\frac{3}{2};-\frac{3c_1}{2\beta}a^6\right].\label{eq28}
\eea
This result can be rewritten as
\bea
&&a^{\frac{3(1+A)}{2}}\,_{2}F_{1}\left[x,x;1+x;-\frac{B}{C(1+A)}a^{\frac{3(1+A)}{2x}}\right]=\frac{\sqrt{3}}{2}(1+A)C^x\,t,\label{eq29}
\eea
where $A=1$, $B=6c_1$, $C=2\beta$ and $x=\frac{1}{2}$. 

We can also analyze this result (\ref{eq28}) by writing the hypergeometric function as a hypergeometric series \cite{Grad}. Doing this, we see that this series is equal to series of the function  arcsinh and so we can write
\bea
t=\frac{1}{3\sqrt{c_1}}\rm{arcsinh}\left[\sqrt{\frac{3c_1}{2\beta}}a^3\right].
\eea
In this way, the scale factor can be written as
\bea
a(t)=\left(\frac{2\beta}{3c_1}\right)^{\frac{1}{6}}\sinh^{\frac{1}{3}}\left(3\sqrt{c_1}\,t\right).\label{eq31}
\eea

Comparing the result (\ref{eq29}) with the developments in \cite{Ujjal} we observe that they are similar. Can also be seen here that the result for the scale factor, equation (\ref{eq31}), is similar to those obtained in \cite{Costa} and \cite{Julio}.

 Therefore, we conclude that we find for suitable choices of some parameters that the Ricci dark energy model in the Chern-Simons gravity displays the same results obtained from the modified Chaplygin gas, i.e., our results show that there is a correspondence between the Ricci dark energy in gravity Chern-Simons and the modified Chaplygin gas. This correspondence between the Ricci dark energy model and the Chaplygin gas has also already been obtained in other contexts as shown in \cite{Chimento} \cite{Pasqua}. Therefore, in our case, we found a gravity model that exhibits the same behavior displayed by the modified Chaplygin gas.

\section*{Acknowledgments}

This work was supported by the Project FAPEMAT/CNPq No. 685524/2010. J. G. Silva thanks Coordena\c{c}\~ao de Aperfei\c{c}oamento de Pessoal de N\'ivel Superior (CAPES) for financial support.

\end{document}